\documentclass[showkeys,nofootinbib]{revtex4}  % for review and submission
\usepackage{graphicx}  % needed for figures
\usepackage{dcolumn}   % needed for some tables
\usepackage{bm}        % for math
\usepackage{amssymb}   % for math
\usepackage{amsmath}
\usepackage{xcolor}
\usepackage{amsthm}
\usepackage{enumitem}
\usepackage[title,titletoc]{appendix}
\usepackage{etoolbox}

\newcommand{\be}{\begin{equation}}\newcommand{\ee}{\end{equation}}
\newcommand{\bea}{\begin{eqnarray}}\newcommand{\eea}{\end{eqnarray}}
\newcommand{\brr}{\begin{array}}\newcommand{\err}{\end{array}}
\newcommand{\bit}{\begin{itemize}}\newcommand{\eit}{\end{itemize}}
\newcommand{\ben}{\begin{enumerate}}\newcommand{\een}{\end{enumerate}}

\newcommand{\ba}{\begin{array}}
\newcommand{\ea}{\end{array}}

\begin{document}
\title{Gravitational effects on the Heisenberg Uncertainty Principle: a geometric approach}

\author{Jaume Gin\' e\footnote{gine@matematica.udl.cat}$^{\hspace{0.3mm}1}$ and Giuseppe Gaetano Luciano\footnote{gluciano@sa.infn.it}$^{\hspace{0.3mm}2,3}$} \affiliation
{$^1$Departament de Matem\`atica, Universitat de Lleida, Av.~Jaume II, 69, 25001 Lleida, Catalonia, Spain.\\
$^2$Dipartimento di Fisica, Universit\`a di Salerno, Via Giovanni Paolo II, 132 I-84084 Fisciano (SA), Italy.\\
$^3$INFN, Sezione di Napoli, Gruppo collegato di Salerno, Italy.\\
}

\date{\today}
\def\be{\begin{equation}}
\def\ee{\end{equation}}
\def\al{\alpha}
\def\bea{\begin{eqnarray}}
\def\eea{\end{eqnarray}}

\begin{abstract}
The Heisenberg Uncertainty Principle (HUP) limits 
the accuracy in the simultaneous measurements of the
position and momentum variables of any quantum system. This is
known to be true in the context of non-relativistic quantum
mechanics. Based on a semiclassical geometric approach, here
we propose an effective generalization of this principle,
which is well-suited to be extended
to general relativity scenarios as well. We apply our formalism
to Schwarzschild and de Sitter spacetime, showing
that the ensuing uncertainty relations
can be mapped into well-known deformations of the HUP.
We also infer the form of the perturbed metric
that mimics the emergence of a discrete spacetime structure at 
Planck scale, consistently with the predictions
of the Generalized Uncertainty Principle. Finally, we discuss
our results in connection with other approaches
recently appeared in the literature.

\end{abstract}

\vskip -1.0 truecm

\maketitle

\makeatletter
\appto{\appendix}{%
  \@ifstar{\def\thesection{\unskip}\def\theequation@prefix{A.}}%
          {\def\thesection{\Alph {section}}}%
}
\makeatother

\section{Introduction}
The Heisenberg Uncertainty Principle (HUP)
plays a pivotal r\^ole within the framework
of non-relativistic Quantum Mechanics (QM) and can be regarded
as one of the cornerstones of quantum theory.
However, in high energy/large distance regimes, 
it is expected to acquire corrections in order to 
accommodate effects that escape
the domain of QM. For instance,
several models of quantum gravity, such as
String Theory~\cite{ST}, Loop Quantum Gravity~\cite{Rovelli} 
and Doubly-Special Relativity~\cite{AmCam}
predict the emergence of a minimal length at Planck scale,
which is at odds with the standard
Heisenberg's formula. This feature is 
taken into account by generalizing the HUP to the so-called
Generalized Uncertainty Principle (GUP), which
naturally embeds the UV length cutoff into a 
minimal position uncertainty.
On the other hand, in cosmological scenarios it has been
argued that the HUP should be modified
by introducing corrections proportional to
the cosmological constant, which reflects the
existence of a maximum length of the order
of the cosmological horizon~\cite{Mignemi}.
Such kind of modification is referred to as Extended Uncertainty Principle (EUP). Deformations that combine 
both the Generalized and Extended Uncertainty Principles (GEUP)
are also being considered as models for a more
complete picture of the quantum spacetime~\cite{Bambi}. 

While providing an effective description
of QM in extreme regimes, the GUP and HUP
break down Lorentz covariance by introducing
a universal length scale. This poses the problem of 
how to extend
these models to the relativistic theory.
Attempts to tackle this issue have
been carried out in Refs.~\cite{Quesne,Quesne2,Quesne3}
via the introduction of the most general covariant
form of the quadratic GUP algebra in Minkowski spacetime.
However, challenging problems remain open, such
as the derivation of the uncertainty relation (UR)
for a generic (curved) background or, by reversing the perspective,  
the determination of the
metric capable of mimicking a given deformed
UR. In this regard, we mention
that gravitationally induced UR's
in the context of both General Relativity (GR) and
extended theories of gravity have been
recently investigated in Ref.~\cite{PetWag}
by assuming the wave function of the quantum 
system to be confined to a
geodesic ball on a given space-like hypersurface 
whose radius is a measure of the position
uncertainty. Furthermore, a metric-dependent uncertainty
relation based on a proper redefinition of the translation operator
has been exhibited in Ref.~\cite{Costa}. 

Starting from the above premises,
in this work we propose a semiclassical geometric approach
to compute a metric-dependent generalization of the HUP.
Specifically, we first extend the canonical 
UR to Minkowski spacetime through a
suitable redefinition of the scalar product.
Then, we apply our considerations to
curved manifolds endowed with a time-like Killing vector. As particular examples,
we deal with Schwarzschild, perturbed weak-field Schwarzschild 
and de Sitter metrics,
showing how the ensuing uncertainty relations
can be mapped into generalizations of the
HUP which are well-established in the literature.
Letting ourselves be guided by the
phenomenological predictions of GUP,
we also provide some hints towards
understanding how Minkowski metric
should be perturbed at quantum gravity scale
in order to reproduce the emergent
discrete structure of spacetime.

The remainder of the work is organized as follows: in Sec.~\ref{RGHUP}
we set the stage for the generalization of the Heisenberg Uncertainty Principle.
The ensuing relation is applied to some specific metrics in 
Sec.~\ref{App}. Conclusions and outlook are summarized in Sec.~\ref{Conc}.
Throughout all the paper, we set
the (reduced) Planck's constant $\hslash$,
the Newton's gravitational constant $G$,
the speed of light in vacuum $c$ and Boltzmann's constant
$k_\mathrm{B}$ equal to unity.
Furthermore, we use the metric with the
conventional spacelike signature
\be
\label{Minkmet}
\eta_{\mu\nu}=\mathrm{diag}(-1,+1,+1,+1)\,.
\ee
Quantum operators will be distinguished from classical variables 
by the usual symbol $\hat{}$\,\,. However, the same notation for both quantities  
will be employed where no ambiguity arises. 

\section{Generalization of Heisenberg Uncertainty Principle}
\label{RGHUP}
Let us consider a particle of position vector $\textbf{x}\equiv(x,y,z)$
and denote by  $E,\textbf{p}\equiv(p_x,p_y,p_z)$ its energy and
three-momentum.  
According to the standard HUP
applied, for example, along the $x$-direction, it is well-known that 
\be
\label{HUP}
\Delta x \Delta p_x \simeq \frac{1}{2}\,,
\ee
where $\Delta x$ and $\Delta p_x$ are the position
and momentum uncertainties of the particle state, respectively.

By assuming that the uncertainties along the three axes do not affect each other, the above relation can be equally written for the remaining
spatial dimensions, yielding
\be
\label{HUP2}
\Delta y \Delta p_y \simeq \frac{1}{2}\,,\qquad
\Delta z \Delta p_z \simeq \frac{1}{2}\,.
\ee
Likewise, one can introduce an equivalent time-energy
uncertainty relation in the form
\be
\label{TE}
\Delta t \Delta E\simeq \frac{1}{2}\,.
\ee
Following one of the earliest proposed versions of
this relation, here we identify $\Delta E$ and $\Delta t$ with
the uncertainty on the energy measurement 
and the duration of such a measuring process, respectively. 
However, this way of interpreting Eq.~\eqref{TE} is not unique
and several reformulations have
appeared in the literature over the years.
For a more detailed review, one can refer to~\cite{Busch}.

We can now recast the UR's in Eqs.~\eqref{HUP} and~\eqref{HUP2} in a more compact form by introducing
the notation
\begin{eqnarray}
\label{vector}
\Delta\textbf{x}&\equiv&(\Delta x, \Delta y, \Delta z)\,,\\[2mm]
\Delta\textbf{p}&\equiv&(\Delta p_x, \Delta p_y, \Delta p_z)\,.
\end{eqnarray}
This leads to
\be
\label{vecHUP}
\Delta\textbf{x}\cdot \Delta\textbf{p}=\Delta x \Delta p_x +
\Delta y \Delta p_y + \Delta z \Delta p_z \simeq \frac{3}{2}\,.
\ee
Here we have used the symbol `` $\cdot$ '' to denote
the scalar product between three-vectors. Notice that a
similar multi-dimensional generalization of HUP has been discussed in Ref.~\cite{Book}.

Equation~\eqref{vecHUP} provides the starting point of our analysis.
Indeed, one can naturally extend it to the four-dimensional Minkowski
spacetime by implementing the following replacement
\be
\label{SR}
\Delta\textbf{x}\cdot \Delta\textbf{p}\,\,\rightarrow\,\,|\Delta x^{\mu}\langle\hat\eta\rangle_{\mu\nu} \Delta p^{\nu}|\,,%\equiv \Delta x^{\mu} \Delta p_{\mu} \,,
\ee
where $\langle\hat \eta\rangle_{\mu\nu}$ is defined as 
the tensor whose generic $(\alpha,\beta)$-element ($\alpha,\beta=\{0,1,2,3\}$)
is the $(\alpha,\beta)$-element of $\eta_{\mu\nu}$ 
evaluated on the expectation value\footnote{Notice that in our quantum picture the presence of the expectation value
in Eq.~\eqref{SR} follows from the fact that 
the metric $\hat \eta_{\mu\nu}$
must be regarded as being composed by operators (rather than classical variables) acting on the state of the quantum system.} 
in the quantum state of the system, i.e.
\be
\label{9}
\langle\hat \eta(\hat x)\rangle_{\alpha\beta}\,\equiv\, \eta_{\alpha\beta}(\langle\hat x\rangle)\,.
\ee
The presence of the absolute value in the definition~\eqref{SR} follows from the fact that the metric tensor contains in general negative terms.
Clearly,
for the specific case of Minkowski metric, we simply have
$\langle\hat\eta\rangle_{\mu\nu}=\eta_{\mu\nu}$.
Furthermore, we have defined
\begin{eqnarray}
\label{4x}
\Delta x^{\mu}&=&(\Delta t, \Delta x, \Delta y, \Delta z)\,,\\[2mm]
\Delta p^{\nu}&=&({\Delta E}, \Delta p_x, \Delta p_y, \Delta p_z)\,.
\label{4p}
\end{eqnarray}

In the above setting, by plugging Eqs.~\eqref{9}-\eqref{4p} into \eqref{SR},
we obtain
\be
\label{HUPrel}
%\Delta x^{\mu} \Delta p_{\mu}
|\Delta x^{\mu}\eta_{\mu\nu} \Delta p^{\nu}|=|-\Delta t \Delta E + \Delta x \Delta p_x +
\Delta y \Delta p_y + \Delta z \Delta p_z| \simeq 1\,.
\ee
It is worth noting that the one-dimensional UR's~\eqref{HUP}-\eqref{TE} can be
recovered by properly restricting the
dimension of the metric tensor
and assuming uncertainties along 
different directions to be independent from each other. For instance, 
with reference to the $\alpha^{th}$-component, $\eta_{\mu\nu}$ in Eq.~\eqref{SR} must be replaced
by its $(\alpha,\alpha)$-element. This straightforwardly gives
\be
\label{11}
|\Delta x_\alpha\,\eta_{\alpha\hspace{0.2mm}\alpha}\,\Delta p_\alpha|=\Delta x_\alpha\,\Delta p_\alpha\simeq \frac{1}{2}\,,\qquad\, \alpha=\{0,1,2,3\}\,.
\ee
We stress that the above relation
must not be intended as summed over $\alpha$. Furthermore, 
the r.h.s. has been fixed by properly taking into account
the dimensional restriction. Alternatively, one can 
 obtain the usual UR's in Minkowski spacetime by using the
veirbein formalism, i.e. by projecting the commutator and the metric tensor on the
tangent space.
%Before proceeding further, a comment is in order here:
%unlike non-relativistic quantum mechanics, it is clear that
%in the relativistic framework the Heisenberg principle
%turns out to be somehow dependent on the choice of the
%reference frame, due to the fact that lengths and momenta
%are not invariant under coordinate transformations. For instance,
%by considering a boost with velocity $v$ along the $x$-direction,
%we have $\Delta x'=\Delta x/\gamma$ and $\Delta t'=\Delta t'\gamma$
%(with $\Delta y$ and $\Delta z$ left unchanged),
%where we have labeled the transformed coordinates with a prime
%and $\gamma=1/\sqrt{1-v^2/c^2}$ is the usual Lorentz factor.
%Similar considerations hold true for the component
%of $\Delta p^{\mu}$. Therefore, Eq.~\eqref{HUPrel}
%gets modified by the appearance of an extra $\gamma$ factor,

Now, the most direct way to 
extend the relation~\eqref{HUPrel} to a generic
spacetime of diagonal metric $g_{\mu\nu}$ endowed with a time-like Killing vector reads\footnote{The generalization of our formalism to non-diagonal metrics, as well as to metric that do not admit a time-like Killing vector is not immediate and will be discussed elsewhere.}
\be
\label{HUPgr}
\langle\hat\eta\rangle_{\mu\nu} \,\rightarrow\, \langle\hat g\rangle_{\mu\nu}\hspace{1mm} \Longrightarrow\hspace{1mm}
|\Delta x^{\mu}\langle\hat g\rangle_{\mu\nu}\ \Delta p^{\nu}|\simeq 1\,,
\ee
where the last equality follows from the fact
that Eq.~\eqref{HUPrel} must be recovered
when going back to Minkowski spacetime\footnote{More generally, the r.h.s. of Eq.~\eqref{HUPgr} may depend on some invariants of the metric, in such a way that, for $g_{\mu\nu}\rightarrow\eta_{\mu\nu}$, Eq.~\eqref{HUPgr} is recovered in its current form. In the absence of a  definitive way
to fix these extra-terms, in what follows we stick to the simplest generalization~\eqref{HUPgr}, leaving a more in-depth study for future work.}.  
We notice that a similar generalization of Heisenberg relation
to curved spacetime has been proposed in~\cite{Capoz} at level
of the canonical commutator between position and momentum operators. 

Before applying the above formalism to some specific examples, 
we point out that
the connection between extensions of the
uncertainty relation and spacetime metrics has also been investigated in Ref.~\cite{Costa}. In that case a translation operator acting in a space with a diagonal metric is introduced to describe the motion of a particle in a quantum system. As a result, it is shown that the momentum operator and the uncertainty relation acquire a metric-dependent structure.
Furthermore, for any metric expanded up to the second order, 
such a formalism naturally leads to an Extended Uncertainty Principle with a minimum momentum dispersion. 
The present analysis goes in the direction of~\cite{Wagner}, where
a duality between theories yielding generalized uncertainty principles and quantum mechanics on nontrivial momentum space is found.

\section{Applications}
\label{App}
For the sake of concreteness, in this Section we consider some applications
of the formula~\eqref{HUPgr}. We derive
the corrections to the Heisenberg principle arising in 
Schwarzschild, perturbed weak-field Schwarzschild and de Sitter backgrounds.
We also infer the form of the perturbed metric
which is best suited to the description of spacetime fluctuations
at Planck scale according to the predictions of GUP. 

\subsection{Schwarzschild spacetime}
\label{Scsp}
As a first example, let us consider the spacetime geometry around
a Schwarzschild black hole of mass $M$
and Schwarzschild radius $r_s=2M$.
In spherical coordinates $(t,r,\theta,\varphi)$, this is described by the well-known
metric tensor
\be
\label{Schwarz}
g_{\mu\nu}=\mathrm{diag}(-e^{2\alpha(r)}, e^{-2\alpha(r)}, r^2, r^2\sin^2\theta)\,,
\ee
where
\be
\label{alphar}
e^{2\alpha(r)}=1-\frac{r_s}{r}\,,\quad r>r_s\,.
\ee
By introducing the analogues of Eqs.~\eqref{4x} and \eqref{4p}
in spherical coordinates for an observer in Schwarzschild metric,
%\begin{eqnarray}
%\Delta x^{\mu}&=&(\Delta t, \Delta r, \Delta \theta, \Delta \varphi)\,,\\[2mm]
%\Delta p^{\nu}&=&(\Delta E, \Delta p_r, \Delta p_\theta, \Delta p_\varphi)\,,
%\end{eqnarray}
we obtain from Eq.~\eqref{HUPgr}
\be
\label{SchwarzUP}
|\Delta x^{\mu}\langle\hat g\rangle_{\mu\nu}  \Delta p^{\nu}|=|- e^{2\alpha(\langle \hat r	\rangle)}\Delta t \Delta E + e^{-2\alpha(\langle \hat r\rangle)} \Delta r \Delta p_r + \langle \hat r\rangle^2 \Delta \theta \Delta p_\theta + \langle \hat r \sin\hat\theta\rangle^2\Delta \varphi \Delta p_\varphi| \simeq 1\,,
\ee
where we have denoted by $p_r$, $p_\theta$ and $p_\varphi$ the radial
and angular components of the particle momentum, respectively.

By analogy with Eqs.~\eqref{11},
let us see how the uncertainty principle~\eqref{SchwarzUP}
appears for each couple of variables separately.
In particular, %for the time-energy uncertainty
%relation in Schwarzschild black hole metric, we have
%\be
%\label{Schwarztimen}
%\langle e^{2\alpha(r)}\rangle\Delta t \Delta E = \left(1-r_s\langle\frac{1}{ r}\rangle\right) \Delta t \Delta E \ge \frac{\hslash}{2}\,.
%\ee
%and observing that $1/r$ ($r>0$) is a convex function,
%we can write
%\be
%\Delta t \Delta E \ge \frac{\hslash}{2} {\left(1-r_s\langle\frac{1}{r}\rangle\right)}^{-1}
%\ge \frac{\hslash}{2} {\left(1-\frac{r_s}{\langle r\rangle}\right)}^{-1}\,.
%\ee
%For distances $r\gg r_s$, we also expect that $\langle r \rangle\gg r_s$, so that
%we can make the following leading-order approximation
%\be
%\label{TEUR}
%\Delta t \Delta E\gtrsim \frac{\hslash}{2} {\left(1+\frac{r_s}{\langle r\rangle}\right)},%=\frac{\hslash}{2}+\frac{\hslash}{2}\frac{r_s}{\langle r\rangle}\,,
%\ee
%where the second term provides the first-order
%correction due to the gravitational source $M$. Clearly,
%for $M\rightarrow0$, the standard time-energy uncertainty
%relation~\eqref{TE} is recovered, as it should be.
we focus on the uncertainty relation between $r$ and
$p_r$, which takes the form\footnote{Notice that, if we considered the generalized UR~\eqref{HUPgr} in the most common inequality form, then Eq.~\eqref{rdrelat2a} itself can be cast as an inequality by using Jensen's relation $f(\langle \hat r\rangle)\le \langle f(\hat r)\rangle$ for the convex function $f(r)=e^{-2\alpha(r)}$.}
\be
\label{rdrelat2a}
\Delta r \Delta p_r \simeq \frac{1}{2}\left({1-\frac{r_s}{\langle\hat r\rangle}}\right)\,.
%\gtrsim\frac{1}{2}\left(1-\frac{r_s}{\langle r\rangle}\right), %\ge\frac{\hslash}{2}\left(1-\frac{r_s}{\Delta r}\right),
\ee
%where we have used Jensen's inequality
%\begin{eqnarray}
%\label{radrelat}
%\langle e^{-2\alpha(r)}\rangle\Delta r \Delta p_r &=& \langle\left(1-\frac{r_s}{r}\right)^{-1} \rangle\hspace{0.2mm} \Delta r \Delta p_r\ge \left(1-\frac{r_s}{\langle r\rangle}\right)^{-1}\Delta r \Delta p_r\ge\frac{1}{2}\,,
%\Delta r \Delta p_r \ge \frac{\hslash}{2}{\left(1-\langle\frac{r_s}{r}\rangle\right)}=\frac{\hslash}{2}-\frac{\hslash}{2}\langle\frac{r_s}{r}\rangle\,.
%\end{eqnarray}
%\begin{equation}
%\label{J1}
%\langle f(r)\rangle\ge f(\langle r\rangle)\,,
%\end{equation}
%for the convex function $f(r)=e^{-2\alpha r}$. %, while
%the last inequality guarantees that the standard
%HUP is recovered in the limit of $g_{\mu\nu}\rightarrow\eta_{\mu\nu}$. 
% Equation~\eqref{radrelat} can then be recast as
%\be
%\label{radrelat2}
%\Delta r \Delta p_r\ge\frac{1}{2}\left(1-\frac{r_s}{\langle r\rangle}\right).%\ge\frac{\hslash}{2}\left(1-\frac{r_s}{\Delta r}\right),
%\ee
%where the last inequality arises from the fact
%that the uncertainty $\Delta r$ is typically much smaller than
%$\langle r \rangle$.
Quite unexpectedly, from this relation it follows that
the closer the particle gets to the
Schwarzschild horizon, the more classical its behavior becomes, since the r.h.s. approaches zero (see Fig.~\ref{figure1}). A possible interpretation for this effect
is provided below in connection with the \emph{classicalization}
(i.e. the recovery of a classical behavior) of
gravitational interaction in high energy regimes predicted by some alternative theories of gravity. 

One can now recognize in Eq.~\eqref{rdrelat2a} a GUP-like
deformation of the uncertainty principle. In fact,
as suggested by gedanken experiments involving
the formation of either gravitational instabilities in high energy scatterings of strings~\cite{VenezGrossMende,VenezGrossMende2, VenezGrossMende3, VenezGrossMende4} or micro black holes~\cite{FS},
it is expected that the standard Heisenberg relation gets
non-trivially modified at  Planck scale.
Specifically, by considering a micro black hole of gravitational radius $r_s\equiv r_s(E)=2E$ proportional to the (centre-of-mass) scattering energy $E$, the HUP should take the form~\cite{FS}
\be
\label{GUP}
\Delta r\simeq \frac{1}{2E} + \beta r_s(E)\,.
\ee
We note that this kind of modification was
also proposed in Ref.~\cite{Adler}. Furthermore, it
can be recast in the most common form of quadratic Generalized
Uncertainty Principle
\be
\label{GUP2}
\Delta r \Delta p_r\simeq\frac{1}{2}\left(1+\beta\ell_p^2\Delta p_r^2\right), 
\ee
where $\ell_p$ is the Planck length. 
In principle, the deformation parameter $\beta$
is not fixed by the theory,
although it is generally assumed to be of order unity
in some models of string theory~\cite{VenezGrossMende,VenezGrossMende2,VenezGrossMende3,VenezGrossMende4}.
We also mention that many studies have recently attempted to
constrain $\beta$ by means of different quantum mechanical or field theoretical approaches~\cite{Bound,Bound2,Bound3,Bound4,Bound5,Bound6,Bound7,Bound8,Bound9,Bound10,Bound11,Bound12,Bound13,CGgrav}
(see Ref.~\cite{Review} for a recent review).
Concerning the sign of $\beta$, it is easy to show that
Eq.~\eqref{GUP2} leads to a minimal position uncertainty
of the order of Planck length for $\beta>0$,
while no threshold is predicted for $\beta<0$~\cite{discst,discst2}.

Now, as displayed in Eq.~\eqref{rdrelat2a}, in the
gravitational field of a Schwarzschild black hole, the
limit on the simultaneous measurement of the
position and momentum of a quantum system is lowered
with respect to Heisenberg bound.
From Eqs.~\eqref{GUP} and \eqref{GUP2}, it is evident that
this feature is characteristic of GUP models with $\beta<0$.
Although most of studies and
gedanken experiments on the GUP seem to support
the $\beta>0$ scenario,
the possibility of having
negative $\beta$ has not yet been fully
ruled out. For instance, in Ref.~\cite{Jizba:2009qf} it has been
shown that the GUP with $\beta<0$ would lead to
a description of the Universe with an underlying
crystal lattice-like structure. Further arguments
have been proposed in Refs.~\cite{CGgrav} and~\cite{Ong:2018zqn},
where the $\beta<0$ framework has proved to be consistent with the
corpuscular picture of gravity~\cite{Corp1,Corp2,Corp3} and
the phenomenologically observed
Chandrasekhar limit for white dwarfs,
respectively. Notice also that our result is in line with Ref.~\cite{ScardCas},
where it has been argued that
a spacetime metric is able to reproduce the GUP-deformed
Hawking temperature, provided that the parameter $\beta$ is
assumed to be negative. Remarkably, in all these contexts
it has been highlighted that, if $\beta<0$,
there exists a maximum $\Delta p$ such
that $\Delta r \Delta p_r\simeq0$  as one approaches Planck scale.
Contrary to common belief, it thus follows that physics in high energy
regime would look classical again,
similarly as in `t Hooft's approach to deterministic quantum mechanics~\cite{thof,elze}
and in deformed special relativity~\cite{DSR}.
From this perspective, the result~\eqref{rdrelat2a} is
easily explained: since the higher the energy scale,
the more classical the behavior of gravity and vice-versa,
effects of quantum fluctuations are expected
to become increasingly negligible
the closer we get to the black hole horizon. In turns, this 
results into a trivialization of the uncertainty relation 
at Schwarzschild radius.

Now, by exploiting the above formalism, one might
reverse the problem and look for the 
metric that exactly fits the GUP~\eqref{GUP2}.
We expect that the most suitable candidate for this r\^ole
is a solution of the would-be quantum reformulation of GR, which must include among
its distinctive features the emergence of a discrete
structure of the spacetime at Planck scale.
In the absence of a consistent theory of quantum gravity,
we then resort to an effective semiclassical
description. In particular, by following the approach of Ref.~\cite{Goklu,Camacho},
we regard the spacetime as a classical background over which quantum Planck scale fluctuations are
imposed. This allows us to write the perturbed metric as
\be
\label{Minkpert}
g_{\mu\nu}=\eta_{\mu\nu}+h_{\mu\nu}\,,
\ee
where, in principle, $h_{\mu\nu}$ can be dependent
on both space and time coordinates. The condition 
$|h_{\mu\nu}|\ll1$ is required to be satisfied.

Let us focus on the modified uncertainty relation
between the radial component of spatial and momentum variables.
By retracing the same steps as in Eq.~\eqref{rdrelat2a},
we are led to
\be
\langle \hat g\rangle_{11}\Delta r \Delta p_r = \left(1+\langle \hat h\rangle_{11} \right)\Delta r \Delta p_r\simeq \frac{1}{2}\,,
\ee
%where we have focused on the $(1,1)$ component of the metric,
%%since this is the only relevant term for the
%computation of the modified uncertainty relation
%between spatial-momentum variables
%we are interested in (see Eq.~\eqref{radrelat}).
%Here $q$ denotes the dimensionless deformation parameter of the metric.
%Notice that Eq.~\eqref{g11} has a singularity for $r=\sqrt{q}\hspace{0.2mm}\ell_p$:
%we require that $q$ is positive and of order unity,
%so that $\ell_p$ effectively acts as a minimal length beyond
%which the metric~\eqref{g11} ceases to be valid.
%Of course, far from the Planck scale, the usual Minkowski framework
%is recovered.
which implies
\be
\label{perturb}
\Delta r \Delta p_r\simeq \frac{1}{2}\left(1+\langle \hat h\rangle_{11}\right)^{-1}\simeq\frac{1}{2}\left(1-\langle \hat h\rangle_{11}\right),
\ee
to the leading order in the perturbation.
By requiring consitency between Eqs.~\eqref{GUP2} and~\eqref{perturb},
we can infer the relation 
\be
\langle \hat h\rangle_{11}\simeq -\beta\ell_p^2\Delta p_r^2=-\beta\ell_p^2\left(
\langle \hat p^2_r\rangle-\langle \hat p_r\rangle^2
\right).
\ee
Clearly, to get positive spacetime fluctuations as discussed in~\cite{SP}, 
we have to set $\beta<0$. 

Now, to the leading order in the deformation parameter $\beta$,
we can safely use the standard HUP and
approximate $\Delta p_r\simeq 1/\left(2\Delta r\right)$, thus yielding
\be
\label{perturbation}
\langle \hat h\rangle_{11}\simeq-\beta\left(\frac{\ell_p}{\Delta r}\right)^2\,,
\ee
where we have reabsorbed the numerical factor
in the definition of $\beta$.
Therefore, based on our
approach, the semiclassical background
consistent with the GUP~\eqref{GUP2}
can be described as a perturbed Minkowski metric,
with a correction given by Eq.~\eqref{perturbation}.
For small perturbations, we can still
require that $g_{00}\simeq -g_{11}^{-1}\simeq -1 +\langle \hat h\rangle_{11}=-1-\beta \ell_p^2/\Delta r^2$.
%In this setting, the metric~\eqref{Minkpert}
%turns out to be singular
%for \textcolor{red}{$\Delta r\simeq|\beta| \ell_p$}  \textcolor{blue}{$\Delta r\simeq \sqrt{-\beta} \, \ell_p$ which implies $\beta <0$} .

As expected, deviations from Minkowski
background become relevant only at Planck scale,
beyond which the form~\eqref{Minkpert} for the perturbed metric
ceases to be valid due to the singularity $\Delta r= \sqrt{|\beta|} \, \ell_p$.
We also emphasize that an alternative possibility
to take into account spacetime discreteness
has been proposed in the context of
conformally-quantized gravity by
considering  fluctuations of the conformal factor only and quantizing them~\cite{Padmanabhan}.
For an almost comprehensive review of
various approaches to the description of
spacetime fluctuations, see Ref.~\cite{Hossen}.

\subsection{Perturbed weak-field Schwarzschild spacetime}
Let us now consider the effects of a nonminimally coupled (NMC)
model of gravity on a perturbed Minkowski metric.
For the description of this model, we basically
follow Ref.~\cite{Castel-Branco:2014exa}.
The metric we use is the one associated with an asymptotically flat spacetime
around a spherical object of mass $M$, radius $r_s$ and static
radial mass density $\rho(r)$. In spherical
coordinates and in the weak-field limit, it is given by the following
perturbation of Minkowski metric
\be
g_{\mu\nu}=\mathrm{diag}\left(-[1+2\Psi(r)], 1+2\Phi(r), r^2, r^2\sin^2\theta\right),
\ee
where $\Psi$ and $\Phi$ are the perturbing
functions such that $|\Psi(r)|\ll1$ and $|\Phi(r)|\ll1$.
Following Ref.~\cite{Castel-Branco:2014exa} and solving the linearized
field equations, it is possible to show that, outside the spherical body ($r> r_s$), these functions are given by
\be
\Psi(r)=-\frac{M}{r}\left[1+\left(\frac{1}{3}-4\xi\right) A(m,r_s)e^{-mr}\right],
\ee
and
\be
\Phi(r)=\frac {M}{r}\left[1-\left(\frac{1}{3}-4\xi\right) A(m,r_s)e^{-mr}(1+mr)\right],
\ee
where $m$ is a characteristic mass scale,
$\xi$ a dimensionless parameter specific of the NMC model
and $A(m,r_s)$ a form factor which can
be found by integrating the field equations of NMC gravity.
Note that, in the GR limit, $\xi=0$ and $m\rightarrow \infty$, so that
the exponential term in both the perturbing functions $\Psi$ and
$\Phi$ vanishes and we recover the weak-field approximation of Schwarzschild metric, as expected.

\begin{figure}[t]
\centering
\begin{minipage}{\columnwidth}
{\resizebox{14cm}{!}{\includegraphics{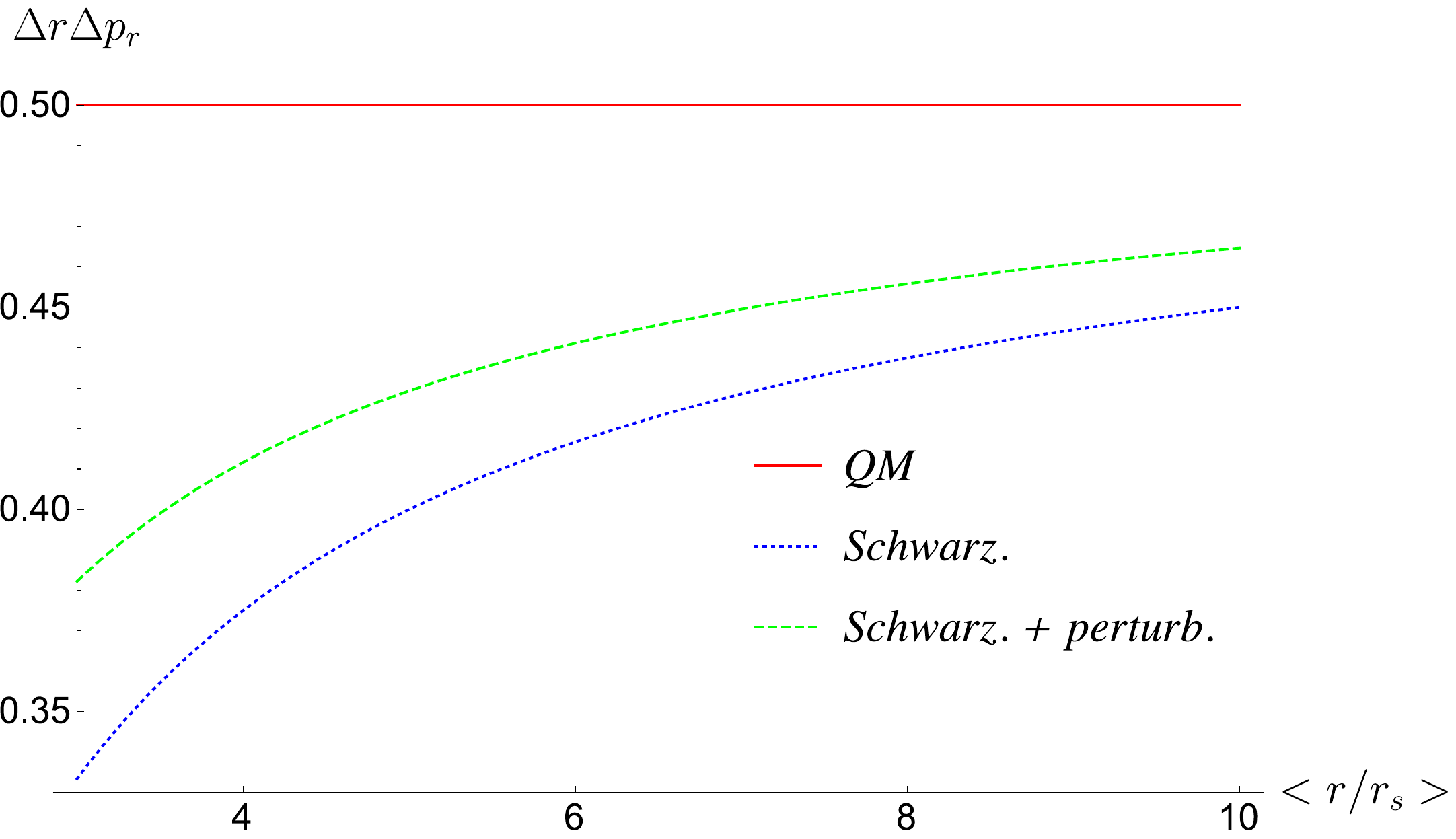}}}
\end{minipage}
\caption{Uncertainty relation for Minkowski (red solid line), Schwarzschild (blue dotted line) and perturbed weak-field Schwarzschild (green dashed line) metrics. For each metric, the allowed region is that above the corresponding curve. For the 
perturbed weak-field Schwarzschild plot, we have set  the sampling values $M=1$, $\xi=10^{-2}$, $m=10^{-4}$ as in Ref.~\cite{Castel-Branco:2014exa}. Notice that, far enough from the source, both the blue-dotted and green-dashed lines converge to the standard Heisenberg limit, as expected.}
\label{figure1}
\end{figure}

As shown in Ref.~\cite{Castel-Branco:2014exa}, the expression
of $A(m,r_s)$ strongly depends on the  mass
density $\rho(r)$. However, by considering a uniform density profile and
assuming not to be too far from the gravity source, we can approximate
$A(m,r_s)\approx1$.
%assuming a uniform density profile, one has
%a form factor with the following asymptotic behavior
%\begin{eqnarray}
%A(m,r_s)&\approx& 1, \quad \mathrm{for}\,\,\,\, m r_s\ll1,\\[2mm]
%A(m,r_s)&\approx& \frac{3}{2} \frac{e^{m r_s}}{(m r_s)^2}, \quad \mathrm{for}\,\,\,\, m r_s\gg1.
%\end{eqnarray}
In this setting, by resorting to
Eq.~\eqref{HUPgr}, the modified uncertainty relation 
between $r$ and $p_r$ in the weak-field limit becomes
%\be
%\left(1+2\langle \Phi(r) \rangle\right) \Delta r \Delta p_r \gtrsim \left(1+2\Phi(\langle r\rangle)\right) \Delta r \Delta p_r \ge \frac{1}{2}\,.
%\ee
\be
\Delta r \Delta p_r \simeq \frac{1}{2}\left(1+2\Phi\left(\langle \hat r\rangle\right)\right)^{-1}\,.%\simeq\frac{1}{2}\left(1-2\Phi(\langle r\rangle)\right)\,,
\ee
From the above equation, we obtain
\begin{eqnarray}
\nonumber
\Delta r \Delta p_r %&\simeq& \frac{1}{2}\left(1+2\Phi(\langle \hat r\rangle)\right)^{-1}
&\simeq&\frac{1}{2}\left(1-2\Phi(\langle \hat r\rangle)\right)
\\[2mm]
&=&\frac{1}{2}\left\{1-2\frac{M}{\langle \hat r\rangle}\left[1-\left(\frac{1}{3}-4\xi\right)e^{-m\langle \hat r\rangle}(1+m\langle \hat r\rangle)\right]\right\},
\label{perbo}
\end{eqnarray}
where in the first step we have expanded
around $|\Phi(r)|\ll1$. As a result, one can see that the quantum gravitational threshold on the simultaneous measurement of $r$ and $p_r$
is slightly increased with respect to the bound in Eq.~\eqref{rdrelat2a} (see Fig.~\ref{figure1}). However, the latter is recovered for $\xi=0$ and $m\rightarrow \infty$, as discussed above. We also point out that both bounds
in Eqs.~\eqref{rdrelat2a} and~\eqref{perbo} 
converge to the standard Heisenberg limit far enough from the source.

In passing, we mention that 
a similar connection between the deformed
Schwarzschild metric and modified uncertainty
relations has been investigated
in Ref.~\cite{Kanazawa:2019llj} within the framework
of  non-commutative geometry~\cite{non-com,non-com2,non-com3}. In that case,  the authors
relate the deformation parameter of non-commutative geometry
to the $\beta$ coefficient appearing in the GUP via the computation of the modified Hawking temperature. Even in that
context a negative value for $\beta$ is obtained, 
thus suggesting a granular structure of spacetime at
the Planck scale.

\subsection{de Sitter spacetime}
\label{dSsp}
As a further application of the above formalism, let us now consider
the case of the four-dimensional de Sitter spacetime.
It is well-known that the main use of de Sitter space
in General Relativity is to describe a simple mathematical
model of the Universe consistent with
its observed accelerating expansion. More
specifically, de Sitter metric represents the maximally symmetric
solution of Einstein's field equations in vacuum
with a positive cosmological constant, corresponding
to a positive vacuum energy density and negative pressure. 

In static coordinates $(t,r,\theta,\varphi)$, de Sitter metric
takes the form
\be
\label{dS}
g_{\mu\nu}=\mathrm{diag}(-e^{2\beta(r)}, e^{-2\beta(r)}, r^2, r^2\sin^2\theta)\,,
\ee
where 
\be
\label{betar}
e^{2\beta(r)}=1-\frac{r^2}{l_H^2}\,, \quad r<l_H.
\ee
Here $l_H$ is the de Sitter radius, which can be expressed in terms
of the cosmological constant $\Lambda$ as
$l^2_H\simeq1/\Lambda$, with $\Lambda>0$.

The computation of the
uncertainty relations for the metric~\eqref{dS}
closely follows the one carried out for
Schwarzschild spacetime. Indeed, with the particular choice
of coordinates we have adopted, the metric tensors~\eqref{dS}
and~\eqref{Schwarz} can be mapped into each other by making the substitution
$e^{2\alpha(r)}\rightarrow e^{2\beta(r)}$.
In this way, we obtain
\be
|\Delta x^{\mu}\langle\hat g\rangle_{\mu\nu} \Delta p^{\nu}|=|- e^{2\beta(\langle\hat r\rangle)}\Delta t \Delta E + e^{-2\beta(\langle \hat r\rangle)}\Delta r \Delta p_r + \langle \hat r\rangle^2 \Delta \theta \Delta p_\theta + \langle \hat r \sin\hat\theta\rangle^2\Delta \varphi \Delta p_\varphi| \simeq 1\,.
\ee
Again, we focus on the uncertainty relation
between $r$ and $p_r$, which can be now rearranged as
%\begin{equation}
%\label{dSUP}
%\langle e^{-2\beta(r)}\rangle\Delta r \Delta p_r = \langle{\left(1-\frac{r^2}{l_H^2}\right)}^{-1}\rangle\Delta r \Delta p_r \ge {\left(1-\frac{\langle r\rangle^2}{l_H^2}\right)}^{-1}\Delta r \Delta p_r\ge{\left(1-\frac{\Delta r^2}{l_H^2}\right)}^{-1}\Delta r \Delta p_r \ge \frac{1}{2}\,,
%\ee
\be
\Delta r \Delta p_r \simeq \frac{1}{2}\left(1-\frac{\langle \hat r\rangle^2}{l_H^2}\right).
\label{dS1bis}
\ee
It is interesting to observe that this equation 
is comparable with the 
%where we have assumed $\Delta r$
%is at most of the order of the mean value of $r$, i.e.
%$\Delta r\le \langle r \rangle$.
%By using the standard relation $\Delta r^2=\langle r^2\rangle-\langle r\rangle^2$,
%this can be rearranged as
%\be
%\label{dS2}
%\Delta r \Delta p_r \ge \frac{\hslash}{2}\left(1+\frac{\Delta r^2}{l_H^2} + \gamma\right),
%\ee
%where $\gamma=-\frac{\langle r\rangle^2}{l_H^2}$.
well-known deformation of HUP arising at cosmic scale,
here rewritten as in~\cite{Gine:2020izd}
\be
\label{EUP}
\Delta r\Delta p \,\simeq\, \frac{1}{2}\left[1+\frac{\gamma}{2}\left(\frac{\Delta r}{l_H}\right)^2\right],
\ee
provided that one assumes $\langle \hat r\rangle \simeq \Delta r$, 
which is quite reasonable at cosmological-scale distances.

The generalization~\eqref{EUP} is usually referred to as
Extended Uncertainty Principle. In this case the r\^ole of the
deformation parameter is
played by the $\gamma$ coefficient. Of course, the
matching between Eqs.~\eqref{dS1bis} and \eqref{EUP}
is met for $\gamma<0$. This finds confirmation
in the analysis of Refs.~\cite{,Mignemi} and more recently of~\cite{Gine:2020izd}, 
where the value $\gamma=-1/\pi^2<0$ has been obtained
by requiring consistency between deformations
of HUP at cosmic scales in de Sitter space and predictions by Modified Newtonian dynamics (MoND) theories. In this regard, we emphasize
that, while for $\gamma>0$ Eq.~(\ref{EUP})
implies the existence of a minimal momentum $p_{min}\sim\sqrt{\gamma}/l_H$, the $\gamma<0$ framework
is characterized only by a maximum length
given of course by the radius $l_H$ of the cosmological horizon~\cite{Mignemi}. 
Clearly, for $\gamma=0$ and/or $\Delta r/l_H\ll 1$,
the standard HUP is recovered. 

Phenomenological implications of the EUP have been
considered in a variety of contexts, ranging
from black hole physics~\cite{Park,Zhu:2008cg,Mureika,Dabrowski,Chung},
to the thermodynamics of the FRW universe~\cite{Zhu:2008cg} and Unruh effect~\cite{Chung}.

On the other hand, one can repeat the same calculation as above
by considering anti-de Sitter spacetime as background metric. 
As opposed to de-Sitter, this metric describes
a Universe with a negative cosmological constant
corresponding to a slowed down expansion. In this case, it is straightforward to
see that the opposite condition $\gamma>0$ is obtained.

Let us finally observe that the results obtained
for the Schwarzschild and de Sitter spacetime
can be merged by looking at the
structure of the uncertainty principle in
de Sitter-Schwarzschild metric.
This metric describes a spherically symmetric solution
with a positive cosmological constant,
providing a spacetime background with
both event and cosmological horizons.
In this framework, the metric tensor can be written as
\be
g_{\mu\nu}=\mathrm{diag}(-e^{2\omega(r)}, e^{-2\omega(r)}, r^2, r^2\sin^2\theta)\,,
\ee
with
\be
e^{2\omega(r)}=1-\frac{r_s}{r}-\frac{r^2}{l_H^2}\,, \quad r_s<r<l_H.
\ee
Following the same considerations as above, one
can see that, far enough from both the Schwarzschild and cosmological
horizon radii, the ensuing HUP
resembles the so-called Generalized Extended Uncertainty Principle (GEUP)~\cite{Adler,Bambi}, which combines
together the effects of the GUP and EUP models.

\section{Conclusions and Outlook}
\label{Conc}
The generalizations of the Heisenberg Uncertainty Principle deduced in the last decades from quantum gravity and
cosmological models have been mimicked starting from well-known
metric solutions of General Relativity and beyond.
This result has been achieved by extending the HUP
on the basis of a semiclassical geometric approach.
The cases of Schwarzschild and de Sitter spacetime
have been studied in detail, showing
that they lead to deformations of the uncertainty relation
which are formally similar to the Generalized and Extended Uncertainty Principles,
respectively. Furthermore, by requiring consistency with
the most common form of GUP predicted by the
string theory, we have inferred the would-be perturbed
Minkowski metric that accounts for the emergence of
a discrete spacetime at Planck scale.
In this framework, we have argued
that the characteristic deformation parameter 
must be negative. This is in line
with the result commonly found
in the literature, i.e. that  a negative $\beta$
typically arises in non-trivial space-time 
having an emergent reticular nature~\cite{discst,discst2}.
A similar argument has also been exhibited in Ref.~\cite{Jizba:2009qf}
by showing that the GUP with $\beta<0$ would be consistent
with a description of the Universe with an underlying
crystal lattice-like structure.

Let us emphasize that semiclassical attempts to establish
a connection between geometric properties
of spacetime and uncertainty relations
are not completely novel in the literature. For instance,
a derivation of GUP from Quantum Geometry has been proposed
in Ref.~\cite{Capoz} in Caianiello's
theory of maximal acceleration. In that case, 
gravitational effects are directly implemented
through a generalization of the canonical commutator
between position and momentum operators, in such a way
that the quadratic GUP is recovered in the framework of
Quantum Geometry theory.  A careful investigation of the relation
between our extension and that proposed in~\cite{Capoz} deserves
more attention. 

On the other hand, 
in Ref.~\cite{PetWag} the influence of space-time geometry on the uncertainty relation has been investigated by assuming
the quantum wave function of the system to be confined
to a geodesic ball of radius proportional to the 
position uncertainty and defining a hermitian momentum operator
that complies with the canonical commutation relations
in the non-relativistic limit of the $3+1$ formalism.
Computations have been developed for some metrics
arising in the context of both General Relativity and
extended theories of gravity, showing that there might
be a direct link between deformations of the
Heisenberg Uncertainty Principle and the 
curvature in energy-momentum space. 
In light of this result, it would be interesting to see
how such a formalism interfaces with
our geometric generalization of HUP.
Another challenging perspective would be the investigation of
our formalism in the context of Banados-Teitelboim-Zanelli black holes 
in light of the outcome of Ref.~\cite{Iorio}.

Besides the above issues, some other aspects remain to be addressed.
Indeed, the full understanding of how to 
deal with deformed uncertainty relations in the relativistic framework
provides an essential element 
to explore the phenomenological implications of a minimal/maximal length scale in Quantum Field Theory, where a  systematic
treatment of the problem is still missing.
In this sense, the present analysis should be intended
as a further step toward this goal. 
Work along this and other directions is presently under active investigation and will be elaborated elsewhere.

\end{document}